\documentstyle[aps,epsf,prl,twocolumn]{revtex}
\begin{document}
\title{Single-Spin Measurement and Decoherence in Magnetic Resonance 
Force Microscopy}
\author{G.P. Berman$^1$, F. Borgonovi$^{1,2}$, H.S. Goan$^{3}$,  
S.A. Gurvitz$^{4}$,  
and V.I. Tsifrinovich$^5$}
\address{$^1$Theoretical Division and CNLS,
Los Alamos National Laboratory, Los Alamos, NM 177545}
\address{$^2$Dipartimento di Matematica e Fisica, Universit\`a Cattolica,
via Musei 41 , 25121 Brescia, Italy,
and I.N.F.M., Unit\`a di Brescia, Italy, and I.N.F.N., sezione
di Pavia , Italy}
\address{$^3$ Center for Quantum Computer Technology, University of New South Wales, Sydney, NSW 2052, 
Queensland, Australia}
\address{$^4$Department of Particle Physics, Weizmann Institute of Science, 
Rehovot 76100, Israel}
\address{$^5$IDS Department, Polytechnic University,
Six Metrotech Center, Brooklyn NY 11201}
\maketitle
\begin{abstract}
We consider a simple version of a cyclic adiabatic inversion (CAI) 
technique in magnetic resonance force microscopy (MRFM). We study the 
problem: What component of the spin is measured in the CAI MRFM? We show that the 
non-destructive detection of the cantilever vibrations provides a measurement 
of the spin component along the effective magnetic field. This result is 
based on numerical simulations of the Hamiltonian dynamics 
(the Schr\"odinger equation) and the numerical 
solution of the master equation. 
\end{abstract}
{PACS:} 03.65.Ta,~03.67.Lx,~76.60.-k \tolerance=10000
\section{Introduction}
Magnetic resonance force microscopy (MRFM) is approaching its ultimate 
goal: a single spin detection \cite{1,2,3}. The most promising approach 
to single spin detection is, probably, cyclic adiabatic inversion (CAI) 
\cite{1}. In this approach, the magnetic moment of the sample changes its 
direction adiabatically following the effective magnetic field. The CAI 
of the spin may act as an ``external force'' driving the resonant 
vibrations of the cantilever or it may affect the frequency of the 
cantilever vibrations driven by another source (e.g. the modern 
``OSCAR'' technique \cite{3}). 

The fundamental question which arises in MRFM single-spin measurement is the 
following: What component of the spin is measured by this technique? Indeed,
 in a simple geometry the cantilever tip oscillating along the $z$-axis 
interacts with the $z$-component of the spin and, consequently, is expected 
to measure the spin $z$-component. From the other side, adiabatic inversion assumes 
that the approximate integral of motion is the 
spin component along the effective magnetic field which rotates in the 
$x-z$ plane. Thus, one might expect that the cantilever measures the spin 
component along the effective magnetic field in the rotating reference frame.

In this work we consider the macroscopic cantilever as itself the measuring device interacting 
with an environment. We assume that the influence of an additional (e.g. optical) device which 
detects the cantilever vibrations is small. This corresponds to the current 
MRFM technique. In Section II we discuss the quantum dynamics of the quasi-classical cantilever 
which describes a generation of the Schr\"odinger cat state associated with two possible 
projections of the spin. In Section III we include an interaction of the cantilever 
with an environment inherent to any measurement processes. The latter leads to  
the decoherence of the two possible trajectories of the cantilever due to interaction 
with the environment. 

\section{Hamiltonian Dynamics}

We considering the simple setup shown in Fig. 1. 


The ferromagnetic particle
 with a magnetic moment $\vec{m}$ is mounted on the cantilever tip. 
The permanent magnetic field, $B_0$, points in the positive $z$-direction.
 A rotating {\it rf} field in the $x-y$ plane,
 $\vec{B}_1\sim\exp[i(\omega t-\varphi(t))]$, is resonant with the spin precession around the $z$-axis. The frequency modulation 
of $\vec{B}_1$ causes the CAI of the spin. Under resonant 
conditions, when the period of the cantilever vibrations matches 
the period of the CAI, the amplitude of the cantilever vibrations 
is expected to increase providing the detection of the spin.

The quantum Hamiltonian of the system in the rotating frame (
in terms of dimensionless parameters) can be written as
$$
{\cal H}=(p^2_z+z^2)/2+\dot\varphi(\tau)S_z-\epsilon S_x-2\eta zS_z.
\eqno(1)
$$
Here 
$$
p_z=P_z/P_q,~z=Z/Z_q,\eqno(2)
$$
${\vec S}$ is the electron spin operator,  
$\epsilon=\gamma B_1/\omega_c$, $\dot\varphi=d\varphi/d\tau$, 
$\eta=gF/2F_q$, $P_z$ and $Z$ are the operators of the effective momentum 
and coordinate of the cantilever tip; $\gamma=g\mu_B/\hbar$ is the spin gyromagnetic 
ratio (absolute value); $\omega_c$ is the cantilever frequency; $F$ is 
the magnetic force between the ferromagnetic particle and the spin when the cantilever tip is at the origin ($z=0$). The origin is chosen at the equilibrium position of the cantilever with no spin;  
$\tau=\omega_ct$ is a dimensionless time. The units 
of the coordinate, momentum, and force are given by
$$
Z_q=(\hbar\omega_c/k_c)^{1/2},~P_q=\hbar/Z_q,~F_q=k_cZ_q, \eqno(3)
$$
where $k_c$ is the cantilever spring constant. Note, that we treat an
 electron spin of a paramagnetic atom whose direction is opposite to the 
direction of the atomic magnetic moment. We assume in (1) that the 
transverse magnetic field points in the negative $x$-direction of the 
rotating frame.

With respect to actual ``reading'' devices  we consider a realistic 
scenario for the MRFM technique which involves non-destructive measurements of the amplitude, frequency, and phase of the cantilever vibrations, 
for example by using a fiber-optic  
interferometer operating in the infrared region. We assume that the 
optical detection of cantilever vibrations does not influence significantly 
the cantilever-spin dynamics. (In practice, this means that the disturbance 
caused by the optical radiation is smaller than the thermal noise 
of the cantilever.) 

In this section we do not consider the interaction with the environment which provides the measurement itself. (See also \cite{cat}.) Thus, we use the Schr\"odinger equation,
$$
i\dot\Psi={\cal H}\Psi,\eqno(4)
$$
for computer simulations of the cantilever-spin dynamics. 
In the $z-S_s$-representation, the wave function, $\Psi$, is a spinor. It contains two components, $\Psi(z,1/2,\tau)$ and $\Psi(z,-1/2,\tau)$, which correspond to the two possible values of $S_z$.
Using the expansion over the eigenfunctions, $u_n$, of 
the oscillator Hamiltonian, $(p_z^2+z^2)/2$, we write these two 
components of the cantilever-spin wave function in the form, 
$$
 \Psi(z,1/2,\tau)=\sum_{n=0}^\infty A_n(\tau)u_n,~ 
\Psi(z,-1/2,\tau)=\sum_{n=0}^\infty B_n(\tau)u_n,\eqno(5)
$$
and derive equations for the amplitudes, $A_n$ and $B_n$,
$$
i\dot A_n=(n+1/2+\dot\varphi/2)A_n-(\eta/\sqrt{2})(\sqrt{n}A_{n-1}+
\sqrt{n+1}A_{n+1})
$$
$$
-(\epsilon/2)B_n,\eqno(6)
$$
$$
i\dot B_n=(n-1/2+\dot\varphi/2)B_n+(\eta/\sqrt{2})(\sqrt{n}B_{n-1}+\sqrt{n+1}
B_{n+1})
$$
$$
-(\epsilon/2)A_n.
$$
The initial conditions describe the quasi-classical state of the cantilever 
tip and a spin which points in the positive $z$-direction, 
$$
A_n(0)=(\alpha^n/\sqrt{n!})\exp(-|\alpha|^2/2),~B_n(0)=0,\eqno(7)
$$
$$
\alpha=[\langle z(0)\rangle+i\langle p_z(0)\rangle]/\sqrt{2}.
$$
In our computer simulations we used the following parameter values,
$$
\eta=0.3,~\epsilon=400,~\dot\varphi=-6000+300\tau~if~\tau\le 20,\eqno(8)
$$
$$
{\rm and}~ 
\dot\varphi=1000\sin(\tau-20),~if~\tau > 20.
$$

The value $\eta=0.3$ can be achieved in current MRFM experiments 
\cite{1,2,3}. The parameters for the transverse magnetic field have 
been chosen to satisfy two conditions: i) the condition of the CAI, 
$|\ddot\varphi|\ll\epsilon^2$, and ii) the effective magnetic field 
produced by the cantilever vibrations on the spin is small in comparison 
with the amplitude of the {\it rf} field: 
$2\eta|\langle z\rangle|\ll\epsilon$. We consider the 
results of the computer simulations reliable if they do not change 
with an increase in the number of basic functions, $u_n$. 

To describe the quasi-classical cantilever, we took the initial average 
energy $\langle E(0)\rangle=|\alpha|^2\gg 1$. The 
number of basic functions, $u_n$, needed to provide reliable results, 
increases with the average energy. So, we can not take 
$|\alpha|$ too big. As we study the driven oscillations of the cantilever, 
our results did not show a significant dependence on the initial conditions. 

The main results of our simulations are the following. The wave function of 
the cantilever-spin system,
which is initially a product of the cantilever and spin parts, quickly 
becomes entangled. The probability distribution to find the 
cantilever at the point $z$ at time $\tau$,
$$
P(z,\tau)=|\Psi(z,1/2,\tau)|^2+|\Psi(z,-1/2,\tau)|^2,\eqno(9)
$$
splits into two peaks, ``big'' and ``small'' 
peaks. (See Fig. 2.) When the peaks are separated, the wave function of the 
cantilever-spin system can be represented as a sum of two spinors,
$$
\Psi(z,s,\tau)=\Psi^{(1)}(z,s,\tau)+\Psi^{(2)}(z,s,\tau),\eqno(10)
$$
where the upper indices ``$1$'' and ``$2$'' refer to the big and the small peaks, 
correspondingly. It was found with the accuracy to 1\% that both spinor wave 
functions, $\Psi^{(k)}(z,s,\tau)$ ($k=1,2$), can be represented as a 
product of the cantilever and spin functions,
$$
\Psi^{(k)}(z,s,\tau)=R^{(k)}(z,\tau)\chi^{(k)}(s,\tau),\eqno(11)
$$
where $\chi^{(1)}(s,\tau)$ describes the spin which points in the 
direction of the external effective field, $(\epsilon,0,-\dot\varphi(\tau))$, 
and $\chi^{(2)}(s,\tau)$ describes the spin which points in the opposite 
direction. The ratio of the probabilities for the 
big and the small peak is determined by
 the initial angle between the external effective magnetic field and the 
spin,
$$
\int|R^{(2)}(z,\tau)|^2dz\Bigg{/}\int|R^{(1)}(z,\tau)|^2dz=\tan^2(\Theta/2),
\eqno(12)
$$
where $\Theta$ is the initial direction of the external effective magnetic 
field ($\tan\Theta =-\epsilon/\dot\varphi(0)=1/15$). If the initial 
conditions describe a spin which points, for example, in the positive 
$x$-direction ($A_n(0)=B_n(0)$), our simulations reveal two peaks with 
approximately equal amplitudes. Thus, the Hamiltonian dynamics clearly 
indicates that the quasi-classical cantilever will measure the spin component 
along the effective magnetic field. Certainly, in the frames of the 
Hamiltonian approach we cannot describe the measurement itself: the 
coherence between the two cantilever peaks does not disappear. 
In other words, the Schr\"odinger equation describes the macroscopic 
Schr\"odinger cat state of the cantilever without effects of decoherence.


\section{Master Equation}

In the previous section, we have presented indications that the cantilever ``measures'' 
the spin component along the direction of the effective magnetic field. 
In this section we describe the measurement process. During the measurement process the coherence between two cantilever trajectories disappears. 
It means that the reduced density-matrix of the cantilever-spin system becomes 
a statistical mixture representing two possible trajectories of the system. 
The main question we are going to answer is the following: Does the cantilever, which interacts with the 
environment, measure the spin component along the effective magnetic field?

To answer this question, we studied the dynamics of the cantilever-spin 
system using the master equation. Our purpose is not just to simulate the expected experiment but rather to present a qualitative verification of the conclusion 
obtained in the previous section. That is why we consider the simplest ``ohmic'' model of the environment in the high-temperature approximation \cite{4}. In this 
approximation the environment is described as an ensemble of harmonic oscillators. The number of oscillators per unit frequency is proportional to the frequency 
in the region below the chosen ``cutoff'' frequency, $\Omega$, and $k_BT\gg\hbar\Omega$. 
The master equation for the density matrix, $\rho$, in the high-temperature 
approximation, is
$$
{\partial\rho_{ss'}(z,z',\tau)\over \partial \tau}=
\Bigg[{i\over 2}\Bigg ({\partial^2\over\partial z^2}
-{\partial^2\over\partial {z'}^2}\Bigg )
-{i\over 2}(z^2-{z'}^2)\eqno(13)
$$
$$
-{\beta\over 2}(z-z')\Bigg(
{\partial\over\partial z}-{\partial\over\partial z'}\Bigg )-D\beta (z-z')^2
$$
$$
-2i\eta(z's'-zs)+i\dot\varphi (s'-s)\bigg ]\rho_{ss'}(z,z',\tau)
$$
$$
-i{\epsilon\over 2}\Bigg [\rho_{s\bar s'}(z,z',\tau)-\rho_{\bar ss'}
(z,z',\tau)\Bigg].
$$
Here  $s,s^\prime=\pm 1/2$, $\bar s=-s$, $\bar s^\prime=-s^\prime$,  $D=k_BT/\hbar\omega_c$, $\beta=1/Q$, where $Q$ is the quality factor 
of the cantilever. Again, we use the expansion over the eigenfunctions, $u_n$, 
$$
\rho_{s,s^\prime}(z,z^\prime,\tau)=\sum_{n,m}A^{s,s^\prime}_{n,m}
(\tau)u_n(z)u^*_m(z^\prime).\eqno(14)
$$
Next, we solve numerically the system of equations for the amplitudes, 
$A^{s,s^\prime}_{n,m}(\tau)$,
 $$
\dot A_{n,m}^{s,s'}(\tau) = \eqno(15)
$$
$$
[i\dot{\varphi}(\tau)(s'-s)+\beta/2
-(n+m+1)D\beta -i(n-m)] \ A_{n,m}^{s,s'}(\tau)-
$$
$$
i\eta s'\sqrt{2m}  A_{n,m-1}^{s,s'}(\tau)- 
i\eta s'\sqrt{2m+2}  A_{n,m+1}^{s,s'}(\tau)+
$$
$$
i\eta s\sqrt{2n}  A_{n-1,m}^{s,s'}(\tau)+
i\eta s\sqrt{2n+2}  A_{n+1,m}^{s,s'}(\tau)+
$$
$$
D\beta\sqrt{m(n+1)}A_{n+1,m-1}^{s,s'}(\tau)+
D\beta\sqrt{n(m+1)}A_{n-1,m+1}^{s,s'}(\tau)+
$$
$$
(D+1/2)\beta\sqrt{(n+1)(m+1)}A_{n+1,m+1}^{s,s'}(\tau)+
$$
$$
(D-1/2)\beta\sqrt{nm}A_{n-1,m-1}^{s,s'}(\tau)-
$$
$$
(D-1/2)\frac{\beta}{2}\sqrt{n(n-1)}A_{n-2,m}^{s,s'}(\tau)-
$$
$$
(D+1/2)\frac{\beta}{2}\sqrt{(n+1)(n+2)}A_{n+2,m}^{s,s'}
(\tau)-
$$
$$
(D-1/2)\frac{\beta}{2}\sqrt{m(m-1)}A_{n,m-2}^{s,s'}(\tau)-
$$
$$
(D+1/2)\frac{\beta}{2}\sqrt{(m+2)(m+1)}A_{n,m+2}^{s,s'}
(\tau)-
$$
$$
i\frac{\epsilon}{2}[ A_{n,m}^{s,-s'}(\tau) - 
A_{n,m}^{-s,s'}(\tau)].
$$

Below we describe the results of our computer simulations for the values 
of parameters in (8). First, setting $\beta=D=0$ we obtain the density 
matrix, $\rho_{s,s^\prime}(z,z^\prime,\tau)$, which exactly corresponds to 
the wave function, $\Psi(z,s,\tau)$, derived from the Schr\"odinger equation.

 
The initial density matrix is represented as a product of the cantilever and 
spin parts,
$$
\rho_{s,s^\prime}(z,z^\prime,0)=\Psi(z,1/2,0)\Psi^*(z^\prime,1/2,0)\left(\matrix{1&0\cr 0&0\cr}\right).
\eqno(16)
$$
The wave function, $\Psi(z,1/2,0)$, describes the quasi-classical state 
of the cantilever,
$$
\Psi(z,0)=\sum_{n=0}^\infty A_n(0)u_n(z).\eqno(17)
$$
The values $A_n(0)$ are given in (7). The initial values, 
$A^{s,s^\prime}_{n,m}(0)$, in (15) can be easily found from (16).

For $\tau>0$, the density matrix describes the entangled state which 
cannot be represented as a product of the cantilever and spin parts. 
The initial peak of $\rho_{s,s^\prime}(z,z^\prime,\tau)$ splits into 
two peaks which are centered along the diagonal $z=z^\prime$, and two 
peaks centered at $z\not=z^\prime$, off the diagonal. The density matrix 
can be represented approximately as a sum of four terms corresponding to four peaks,
$$
\rho_{s,s^\prime}(z,z^\prime,\tau)=\rho^{(1)}_{s,s^\prime}+\rho^{(2)}_{s,s^\prime}+\rho^{(3)}_{s,s^\prime}+
\rho^{(4)}_{s,s^\prime},\eqno(18)
$$
where we omit variables, $z,z^\prime,\tau$. The matrices, 
$\rho^{(1)}$ and $\rho^{(2)}$, describe the ``big'' 
and ``small'' diagonal peaks; $\rho^{(3)}$ and $\rho^{(4)}$, 
describe the peaks   
centered at $z\not=z^\prime$.


As an illustration, 
we show in Fig. 3 the quantity,
$$
|\rho_{1/2,1/2}(z,z^\prime,\tau)+\rho_{-1/2,-1/2}(z,z^\prime,\tau)|.
$$
We have found that with accuracy to 1\% the density matrix, 
$\rho^{(1)}_{s,s^\prime}(z,z^\prime,\tau)$,
can be represented as a product of the coordinate and spin parts,
$$
\rho^{(1)}_{s,s^\prime}(z,z^\prime,\tau)=
\hat R^{(1)}(z,z^\prime,\tau)
\hat\chi^{(1)}_{s,s^\prime}(\tau),\eqno(19)
$$
where $\hat\chi^{(1)}_{s,s^\prime}(\tau)$ describes the 
spin which points in the direction of the external effective magnetic 
field $(\epsilon,0,-\dot\varphi(\tau))$. A similar expression is valid for 
$\rho^{(2)}_{s,s^\prime}(z,z^\prime,\tau)$. But in this case, 
$\hat\chi^{(2)}_{s,s^\prime}(\tau)$ describes a spin which points in the 
opposite direction. 

First we note that in order to describe the measurement process (the decoherence), we have to consider an ensemble of quasi-classical cantilevers with the same initial conditions. 
At the same time, we are considering driven oscillations of the cantilever. 
So, the result of our simulations qualitatively does not depend on the initial conditions of the cantilever. 
Second, as we already mentioned, we are going to verify qualitatively the conclusion derived 
in the previous section rather than simulate the expected experiment. Thus, we choose the values of parameters which help us to save a computational time. Namely, we choose a
relatively small (but still quasi-classical) values for the initial energy of the cantilever, and a relatively small value for the thermal parameter $D$ (without violating the high-temperature 
approximation which requires $D\gg 1$). The small initial energy 
of the cantilever allows us to reduce a number of basis functions, $u_n(z)$. A relatively small value of $D$ allows us to observe four well separated peaks at relatively small values of time, $\tau$.

The initial uncertainty of the cantilever position is, $\delta z=1/\sqrt{2}$.
Due to thermal diffusion, the uncertainty of the cantilever 
position increases with time. Thus, we have two effects: i) the increase of 
the amplitude of the driven cantilever vibrations (similar to the 
Hamiltonian dynamics) and ii) the increase of the uncertainty of the 
cantilever position due to thermal diffusion. If the second effect 
dominates, the two positions of the diagonal peaks (i.e. peaks centered 
on the line $z=z^\prime$) become indistinguishable. In this case, one cannot provide a spin measurement with two possible outcomes. 

We have found that peaks centered on the diagonal retain the main 
properties described by the Hamiltonian dynamics. The density matrix, 
$\rho^{(k)}_{s,s^\prime}(z,z^\prime,\tau)$, for $k=1,2$ can be approximately represented 
as a product of the cantilever and spin parts. The spin part of the matrix 
describes the spin which points in the direction of the external effective 
magnetic field ($k=1$) or in the opposite direction ($k=2$). 

Next, we discuss the two peaks centered at $z\not=z^\prime$. 
As an illustration, Figs. 4 and 5 show the contours of the quantities,
$$
|\rho_{1/2,1/2}(z,z^\prime,\tau)+\rho_{-1/2,-1/2}(z,z^\prime,\tau)|,
$$
and
$$
|\rho_{1/2,-1/2}(z,z^\prime,\tau)+\rho_{-1/2,1/2}(z,z^\prime,\tau)|,
$$
given in logarithmic scale. One can see the peaks 
centered at $z\not=z^\prime$ as well as at $z=z^\prime$. The peaks 
centered at $z\not=z^\prime$ describe the coherence between the two 
cantilever positions. The amplitude of these peaks quickly decreases 
due to the decoherence. Thus, the master equation explicitly describes 
the process of measurement. The coherence between two cantilever 
trajectories (the macroscopic Schr\"odinger cat state) quickly disappears. 
As a result, 
the cantilever will ``choose'' one of two possible trajectories. 
Correspondingly, (depending on the cantilever trajectory) the spin 
will point in the direction of the effective magnetic field or in the
 opposite direction.


\section{Conclusion}

We have studied the quantum dynamics of the cantilever-spin system in a simple 
version of the cyclic adiabatic inversion (CAI) magnetic resonance force 
microscopy (MRFM). In this version, the spin experiences a CAI under the 
action of the external phase modulated {\it rf} magnetic field. If the 
frequency of CAI matches the cantilever frequency, the amplitude of the 
cantilever vibrations increases allowing single-spin 
detection. We have studied the problem: Which component of the spin is 
measured by the cantilever? We argue that one will measure the component 
of the spin along the direction of the effective magnetic field
providing non-destructive 
detection of the cantilever vibrations.  This result was first derived using computer simulations of the Hamiltonian dynamics 
(the Schr\"odinger equation). Then, it was confirmed by the numerical 
solution of the master equation. We have considered the case when the 
amplitude of the driven cantilever vibrations was greater than the thermal 
noise. In this case, the phase of the driven vibrations depends on the spin 
component along the direction of the external effective magnetic field. 
Thus, detecting the phase of the cantilever vibrations one can measure 
the spin component along the effective magnetic field.

We should mention that the direct relation between the cantilever trajectory 
and the direction of the spin has been verified for a transient process in 
the CAI MRFM. Our computer capabilities do not allow us to check this 
relation for the stationary cantilever vibrations at $\tau\gg Q$. Also, 
we completely ignored the direct interaction between the spin and the 
environment. We are now investigating this interaction. 

\begin{figure}
\caption{MRFM setup.}
\label{ff1}
\end{figure}

\begin{figure}
\caption{The probability distribution, $P(z,\tau)$, for the cantilever position,
in the logarithmic scale. The values of parameters are:
 $\epsilon=400$ and $\eta=0.3$. The initial conditions are:
$\langle z(0)\rangle=-20$, $\langle p_z(0)\rangle=0$ (which
corresponds to $\alpha=-10\sqrt{2})$.}
\label{ff2}
\end{figure}

\begin{figure}
\caption{Three-dimensional plot of $\log|\rho_{1/2,1/2}(z,z^\prime,\tau)+
\rho_{-1/2,-1/2}(z,z^\prime,\tau)|$, in the logarithmic scale.
The values of parameters are: $\epsilon=400$, $\eta=0.3$, $\beta=D=0$.
The initial conditions are: $\langle z(0)\rangle=-4$, $\langle p_z(0)
\rangle=0$. The colors in Figs. 3-5 correspond to 
the following values for the logarithm of the density matrix: 
white ($<$-16), red (-16,-12), green (-12,-8), blue (-8,-4), 
yellow ($>$-4).}
\label{ff3}
\end{figure}

\begin{figure}
\caption{The contours for $\log|\rho_{1/2,1/2}(z,z^\prime,\tau)+
\rho_{-1/2,-1/2}(z,z^\prime,\tau)|$. The values of parameters are:
 $\epsilon=400$, $\beta=0.001$, and $D=10$. 
The initial conditions are: $\langle z(0)\rangle=-4$, $\langle p_z(0)\rangle=0$.}
\label{ff4}
\end{figure}

\begin{figure}
\caption{The same as in Fig. 4, but 
for $\log|\rho_{1/2,-1/2}(z,z^\prime,\tau)+
\rho_{-1/2,1/2}(z,z^\prime,\tau)|$.}
\label{ff5}
\end{figure}

\section*{Acknowledgments}

This work was supported by the Department of Energy (DOE)
under the contract W-7405-ENG-36. GPB and VIT  thank 
the National Security Agency (NSA), the Advanced Research and 
Development Activity (ARDA), and the Defense Advanced Research Projects Agency (DARPA)  Program MOSAIC for partial support. 

\end{document}